\documentstyle[amstex,aps,amssymb,epsf,floats,twocolumn]{revtex}
\begin{document}
\title{Optical absorption in the soliton-lattice state \\
of a double-quantum-well system}
\author{R. C\^ot\'e}
\address{D\'{e}partement de physique and Centre de Recherche sur les Propri\'{e}t\'{e}
\'{E}lectroniques de Mat\' {e}riaux Avanc\'{e}s,\\
Universit\'{e} de Sherbrooke, Sherbrooke, Qu\'{e}bec, Canada, J1K 2R1}
\date{\today }
\maketitle

\begin{abstract}
When the separation between layers in a double-quantum-well system is
sufficiently small, the ground state of the two-dimensional electron gas at
filling factor $\nu =1$ has an interwell phase coherence even in the absence
of tunneling. For non-zero tunneling, this coherent state goes through a
commensurate-incommensurate transition as the sample is tilted with respect
to the quantizing magnetic field at $\nu=1$. In this article, we compute the optical
(infrared) absorption spectrum of the coherent state from the commensurate
state at small tilt angle to the soliton-lattice state at larger tilt angle
and comment on the possibility of observing experimentally the distinctive
signature of the soliton lattice.
\end{abstract}

\noindent Pacs: 73.21.Fg, 73.43.Lp, 78.30.Fs

\section{Introduction}

At strong magnetic fields and for sufficiently small separation between the
wells, the two-dimensional electron gas (2DEG) in a double quantum well
system (DQWS) can have a broken symmetry ground state with a non zero
interlayer phase coherence even in the absence of tunneling. At filling
factor $\nu =1$, the quantum Hall effect is observed in such coherent state
when the separation between the wells does not exceed some critical value $%
d_{c}$ above which the coherence is lost and the 2DEG becomes compressible.
In DQWS systems, the interlayer phase coherence gives rise to a rich variety
of quantum and finite-temperature phase transitions as well as to some
exotic topological excitations such as merons and bimerons. Some of these
various phases and excitations are reviewed in details in Refs. \cite{bigmac}%
,\cite{bigmac2}, and \cite{dassarma}.

One convenient way to describe the coherent ground states of the 2DEG in a
DQWS is by using a mapping to an equivalent spin-1/2 system. In these
states, the real spins are assumed completely frozen and, in the pseudospin
language, an electron in the left(right) layer is equated with an up(down)
pseudospin. Quantum mechanics allows for any linear superposition of these
states and the corresponding pseudospin can point in any direction in space.
An interesting phase transition, first reported by Murphy et al.\cite{murphy}%
, occurs when, at filling factor $\nu =1$, the sample is tilted with respect
to the quantizing magnetic field. A study of the behavior of the activation
gap as the sample is tilted shows evidence for a phase transition between
two competing quantum Hall ground states. Yang et al.\cite{yang} explained
this change of behavior as a transition between a commensurate and an
incommensurate ground states. This transition can be briefly described in
the following way. In a suitable gauge, the tunneling amplitude modified by
the parallel component of the magnetic field acts as an effective magnetic
field for the pseudospins. This effective field rotates in space with a
wavevector given by $Q=d/\ell _{||}^{2}$ where $d$ is the separation between
the wells and $\ell _{||}^{2}= \hbar c/eB_{||}$ is the magnetic length for
the parallel magnetic field. For $d\neq 0,$ the pseudospins that are forced
to ly in the plane of the wells to minimize the capacitive energy will
locally align with this effective magnetic field. This is the commensurate
(C) state. As the period of the field increases, however, the gain in
tunneling energy obtained by aligning with the field is opposed by the cost
in exchange energy of having non parallel pseudospins. At a critical field,
the exchange energy exceeds the tunneling energy and the pseudospins cease
to rotate with the effective field, behaving almost as if $t=0.$ But, this
incommensurate (I) state is never the ground state of the system. Above some
critical parallel field or wavevector $Q_{c}$, the C state has defects in
the form of sine-Gordon solitons where the phase of the pseudospin slips by $%
2\pi $. For $Q\geq Q_{c}$, the ground state of the systems is a lattice of
these kinks, a soliton-lattice state (SLS).

Apart from the orignal experiment of Murphy et al. there has been no other
experimental signature of the SLS. On the theoretical side, Hanna et al.\cite%
{hanna} have discussed the possibility of detecting this lattice by surface
acoustic wave technique or by measuring the small contribution of the SLS to
the parallel-field magnetization of the DQWS. Some possible ways of
detecting the pseudospin phase solitons in a DQWS have also been described
by Kyriakidis {\it et al}.\cite{loss} Read\cite{read} has described in
details the behavior of the energy gap near the C-I transition. Some details
of the absorption spectrum but in the absence of a parallel magnetic field
has already been worked out by Joglekar {\it et al}\cite{joglekar}. In this
work, we explore another possible signature of the SLS. Building on earlier
work\cite{cotesls} where we derived the collective excitations of the SLS
ground state, we compute the signature of these collective modes in an
absorption experiment. Our basic idea is the following. The dispersion
relation of the collective mode in the SLS has many branches. The
lowest-energy branch corresponds to the Goldstone mode that restores the
broken translational symmetry. The higher-energy branches have non-zero
frequencies at zero wavector. They (as well as the Goldstone mode) involve
motion of the $z$ component of the pseudospins, a component that is related
to changes in the charge density balance between the two layers. These modes
can be excited by an external electromagnetic wave. A nice feature of the
SLS is that simply changing the magnitude of the parallel magnetic field (%
{\it i.e}. tilting the sample) modifies the period of the lattice and,
consequently, the set of frequencies at zero wavevector. In principle, it
could be possible to track the complete dispersion relation of the
pseudospin collective modes by measuring the absorption as the parallel
magnetic field is changed.

This paper is organized as follow. In section II, we relate the
electromagnetic absorption to the polarisation tensor of the DQWS. In
section III, we derive an expression for this polarisation tensor in terms
of the $z$ component of the pseudospin. Section IV gives a brief review of
the commensurate-incommensurate transition. In section V and VI, we compute
the absorption in the C,I and SL states and comment on the possibility of
experimentally observing the resulting spectrum.

\section{Absorption of light in a DQWS}

The coherent ground states of the 2DEG in a QDWS are characterized by
spatial modulations of one or several of their order parameters. If we
specify a pseudospin configuration by the spherical-coordinate fields $%
\theta \left( x,y\right) $ which describes the difference in charge density
between the layers, and $\varphi \left( x,y\right) ,$ which describes the
relative phase of electrons in the right and left wells, then the SLS\
ground state has spatial modulations in $\varphi \left( x,y\right) $ only
and $\theta \left( x,y\right) =\pi /2$ everywhere. Other states such as a
lattice of bimerons\cite{bimeron} in a DQWS would have spatial modulations
in both the phase and the relative occupation of the two wells and so would
the coherent charge-density-wave recently studied by Brey and Fertig\cite%
{fertigbrey}. The approach we develop can be applied to these later two
cases as well. In all these states, the reponse functions are non local in
space and must be described by tensors of the form $A_{\nu ,\mu }\left( {\bf %
r},{\bf r}^{\prime },\omega \right) $ where $\nu ,\mu =x,y,z$. In this
section, we derive a relation between the electromagnetic absorption and the
polarisation response tensor in the coherent states.

In a stationnary regime, the energy absorbed per unit time in a system is
given by the Joule heating term integrated over all the volume, $V$, of this
system. This energy must also be equal to the difference between the
incident and transmitted or diffused energy which is given by an integral of
the Poynting vector over the surface $S_{V}$ of the sample with outward
normal $\widehat{{\bf n}}$: 
\begin{eqnarray}
P\left( t\right)  &=&\int_{V}\,dV{\bf \,E}\left( {\bf r},t\right) \cdot {\bf %
j}\left( {\bf r},t\right)   \label{un} \\
&=&-\frac{c}{4\pi }\int_{S_{V}}dS\,\left[ {\bf E}\left( {\bf r},t\right)
\times {\bf B}\left( {\bf r},t\right) \right] \cdot \widehat{{\bf n}}. 
\nonumber
\end{eqnarray}%
We remark that the absorbed energy does not, in general, correspond to the
difference between the transmitted and incident energy of the
electromagnetic wave since there will be diffusion of the light in an
heterogeneous state. For harmonically varying current and electric field,
the average power dissipated per unit volume at frequency $\omega $ is given
by 
\begin{eqnarray}
P\left( \omega \right)  &=&\frac{1}{2}%
%TCIMACRO{\func{Re}}%
%BeginExpansion
\mathop{\rm Re}%
%EndExpansion
\left[ \int_{V}\,dV{\bf E}^{\ast }\left( {\bf r},\omega \right) \cdot {\bf j}%
\left( {\bf r},\omega \right) {\bf \,}\right]   \label{deux} \\
&=&\frac{1}{2V}\sum_{{\bf q}}%
%TCIMACRO{\func{Re}}%
%BeginExpansion
\mathop{\rm Re}%
%EndExpansion
\left[ \,{\bf E}^{\ast }\left( {\bf q},\omega \right) \cdot {\bf j}\left( 
{\bf q},\omega \right) \right] ,  \nonumber
\end{eqnarray}%
%\begin{equation}
%P\left( \omega \right) =\frac{1}{2}%
%TCIMACRO{\func{Re}}%
%BeginExpansion
%\mathop{\rm Re}%
%EndExpansion
%\left[ \int_{V}\,dV{\bf E}^{\ast }\left( {\bf r},\omega \right) \cdot {\bf j}%
%\left( {\bf r},\omega \right) {\bf \,}\right] 
%=\frac{1}{2V}\sum_{{\bf q}}%
%TCIMACRO{\func{Re}}%
%BeginExpansion
%\mathop{\rm Re}%
%EndExpansion
%\left[ \,{\bf E}^{\ast }\left( {\bf q},\omega \right) \cdot {\bf j}\left( 
%{\bf q},\omega \right) \right] ,  \label{deux}
%\end{equation}%
where we have defined the Fourier transfrom of the current and of the
electric field by 
\begin{eqnarray}
{\bf j}\left( {\bf r},\omega \right)  &=&\frac{1}{V}\sum_{{\bf q}}{\bf j}%
\left( {\bf q},\omega \right) e^{i{\bf q}\cdot {\bf r}},  \label{trois} \\
{\bf E}\left( {\bf r},\omega \right)  &=&\frac{1}{V}\sum_{{\bf q}}{\bf E}%
\left( {\bf q},\omega \right) e^{i{\bf q}\cdot {\bf r}}.  \label{quatre}
\end{eqnarray}%
In Eq. (\ref{deux}), ${\bf E}$ is the electric field {\it in} the sample.

We can relate $P\left( \omega \right) $ to the conductivity tensor by using
the general relation between conductivity and electric field 
\begin{equation}
{\bf j}\left( {\bf q},\omega \right) =\frac{1}{V}\sum_{{\bf q}^{\prime }}%
{\bf \sigma }\left( {\bf q},{\bf q}^{\prime },\omega \right) \cdot {\bf E}%
\left( {\bf q}^{\prime },\omega \right) .  \label{cinq}
\end{equation}
Alternatively, we can also formaly relate the current to the external field
by 
\begin{equation}
{\bf j}\left( {\bf q},\omega \right) =\frac{1}{V}\sum_{{\bf q}^{\prime }}%
\widetilde{{\bf \sigma }}\left( {\bf q},{\bf q}^{\prime },\omega \right)
\cdot {\bf E}_{e}\left( {\bf q}^{\prime },\omega \right) ,  \label{six}
\end{equation}
where, in contrast to ${\bf \sigma ,}$ the tensor $\widetilde{{\bf \sigma }}$
is directly related to the full current-current response function (see, for
example Ref. \cite{wallis}). ${\bf \sigma }$ can be considered as an
unscreened response function, a response to the total electric field that
includes screening corrections, while $\widetilde{{\bf \sigma }}$ is the
screened response to the bare electric field. We have defined the Fourier
transform of the conductivity tensor by 
\begin{equation}
{\bf \sigma }\left( {\bf r},{\bf r}^{\prime },\omega \right) =\frac{1}{V^{2}}%
\sum_{{\bf q},{\bf q}^{\prime }}e^{i{\bf q}\cdot {\bf r-q}^{\prime }\cdot 
{\bf r}^{\prime }}{\bf \sigma }\left( {\bf q},{\bf q}^{\prime },\omega
\right) .
\end{equation}

With Eq. (\ref{six}), the absorbed power is then 
\begin{equation}
P\left( \omega \right) =\frac{1}{2V^{2}}\sum_{{\bf q,q}^{\prime }}%
%TCIMACRO{\func{Re}}%
%BeginExpansion
\mathop{\rm Re}%
%EndExpansion
\left[ \,{\bf E}^{\ast }\left( {\bf q},\omega \right) \cdot \widetilde{{\bf %
\sigma }}\left( {\bf q},{\bf q}^{\prime },\omega \right) \cdot {\bf E}%
_{e}\left( {\bf q}^{\prime },\omega \right) \right] .  \label{sixb}
\end{equation}%
Maxwell's equations can be used to relate the total and external electric
fields 
\begin{align}
{\bf E}\left( {\bf q},\omega \right) & ={\bf E}_{e}\left( {\bf q},\omega
\right)   \label{sept} \\
& -\left( \frac{4\pi i}{\omega }\right) \frac{1}{V}\sum_{{\bf q}^{\prime
}}K\left( {\bf q},\omega \right) \cdot \widetilde{{\bf \sigma }}\left( {\bf q%
},{\bf q}^{\prime },\omega \right) \cdot {\bf E}_{e}\left( {\bf q}^{\prime
},\omega \right) ,  \nonumber
\end{align}%
where the tensor $K\left( {\bf q},\omega \right) $ is defined by 
\begin{equation}
K\left( {\bf q},\omega \right) =\widehat{{\bf q}}\widehat{{\bf q}}{\bf +}%
\frac{1}{1-c^{2}q^{2}/\omega ^{2}}\left( {\bf I}-\widehat{{\bf q}}\widehat{%
{\bf q}}\right) .
\end{equation}%
Inserting Eq. (\ref{sept}) in Eq. (\ref{sixb}), we get (one can show that
the second term in Eq. (\ref{sept}) does not contribute to the real part of
the expression in Eq. (\ref{sixb})) 
\begin{equation}
P\left( \omega \right) =\frac{1}{2V^{2}}\sum_{{\bf q,q}^{\prime }}%
%TCIMACRO{\func{Re}}%
%BeginExpansion
\mathop{\rm Re}%
%EndExpansion
\left[ \,{\bf E}_{e}^{\ast }\left( {\bf q},\omega \right) \cdot \widetilde{%
{\bf \sigma }}\left( {\bf q},{\bf q}^{\prime },\omega \right) \cdot {\bf E}%
_{e}\left( {\bf q}^{\prime },\omega \right) \right] ,
\end{equation}%
an expression that relates the absorbed power to the external electric
field. All local field corrections are included in $\widetilde{{\bf \sigma }}
$. For an external electromagnetic field in the form of a plane wave with
amplitude $E_{0}$, unit polarisation vector ${\bf \xi }$, and wavevector $%
{\bf q}_{0}$, we have 
\begin{equation}
P\left( \omega \right) =\frac{\left| E_{0}\right| ^{2}}{2V^{2}}%
%TCIMACRO{\func{Re}}%
%BeginExpansion
\mathop{\rm Re}%
%EndExpansion
\left[ {\bf \xi }\,\cdot \widetilde{{\bf \sigma }}\left( {\bf q}_{0},{\bf q}%
_{0},\omega \right) \cdot {\bf \xi }\right] .
\end{equation}

Now, from the relation between dielectric and polarisation tensors 
\begin{eqnarray}
{\bf \varepsilon }\left( {\bf q},{\bf q}^{\prime },\omega \right) &=&I\delta
_{{\bf q},{\bf q}^{\prime }}+\frac{4\pi i}{\omega }{\bf \sigma }\left( {\bf q%
},{\bf q}^{\prime },\omega \right) \\
&=&I\delta _{{\bf q},{\bf q}^{\prime }}+4\pi {\bf \chi }\left( {\bf q},{\bf q%
}^{\prime },\omega \right) ,  \nonumber
\end{eqnarray}
we have 
\begin{equation}
{\bf \sigma }\left( {\bf q},{\bf q}^{\prime },\omega \right) =-i\omega {\bf %
\chi }\left( {\bf q},{\bf q}^{\prime },\omega \right) ,
\end{equation}
or alternatively 
\begin{equation}
\widetilde{{\bf \sigma }}\left( {\bf q},{\bf q}^{\prime },\omega \right)
=-i\omega \widetilde{{\bf \chi }}\left( {\bf q},{\bf q}^{\prime },\omega
\right) ,
\end{equation}
where $\widetilde{{\bf \chi }}$ is the response function that takes into
account all local field corrections. The absorption is thus related to the
imaginary part of the polarisation tensor by 
\begin{equation}
P\left( \omega \right) =\frac{\omega \left| E_{0}\right| ^{2}}{2V^{2}}%
%TCIMACRO{\func{Im}}%
%BeginExpansion
\mathop{\rm Im}%
%EndExpansion
\left[ \,{\bf \xi }\cdot \widetilde{{\bf \chi }}\left( {\bf q}_{0},{\bf q}%
_{0},\omega \right) \cdot {\bf \xi }\right] .  \label{abso1}
\end{equation}
For non-interacting electrons, $P\left( \omega \right) $ of Eq. (\ref{abso1}%
) is equivalent to the usual absorption formula given by the Fermi golden
rule.

\section{Polarisation response function of the DQWS}

We consider a symmetric DQWS where $d$ is the separation between the wells
measured from center to center. This DQWS is placed in a strong quantizing
magnetic field directed along the growth axis $z$. The magnetic field can be
tilted from towards the plane of the wells, but its perpendicular component
must be such as to maintain a total filling factor $\nu =1$ in the case of
the SLS. We write the total field as ${\bf B}=B_{||\ }{\bf y}+B_{\perp }{\bf %
z.}$ In the Landau gauge, the vector potential is ${\bf A}=(0,B_{_{\perp }\
}x,{\bf -}B_{||}x)${\bf . }In this work, we consider only the case of an
unbiased DQWS so that the electric charge is equally distributed between the
two layers in the ground state. At low temperature and in the strong
magnetic field limit, we keep only one electric subband in each well and one
Landau level ($n=0$) in the description of the electronic states. The
non-interacting wavefunctions for each well taken separately are given by 
\begin{equation}
\phi _{i,X}({\bf r})=\frac{1}{\sqrt{L_{y}}}\frac{1}{(\pi \ell ^{2})^{1/4}}%
e^{-iXy/\ell ^{2}}e^{-(x-X)^{2}/2\ell ^{2}}\chi _{i}(z),
\end{equation}
where $\ell _{\bot }^{2}=\hbar c/eB_{\perp }$ defines the magnetic length
for the {\it perpendicular} component of the magnetic field and $\chi
_{i}(z) $ with $i=R,L$ are the envelope wave functions of the lowest-energy
states centered on the right or left well. $X$ is the guiding center quantum
number. The degeneracy of each Landau level is given by $N_{\phi }=S/2\pi
\ell _{\bot }^{2}$ where $S$ is the area of the two-dimensional electron
gas. With $N$ electrons in the DQWS, the total filling factor is $\nu
=N/N_{\phi }$. We define another magnetic length associated with the
parallel component of the magnetic field by $\ell _{||}^{2}=\hbar c/eB_{||}$%
. For simplicity, we describe the DQWS in a narrow well approximation {\it %
i.e.} we assume that the width, $b$, of the wells is small ($b<<d$) and
treat interlayer hopping in a tight-binding approximation\cite{cotesls}.

Taking the charge of the electron to be $-e$ and measuring all positions
with respect to the center of the DQWS, the polarisation density operator in
second quantization is 
\begin{equation}
{\bf p}\left( {\bf r},z\right) =-e\sum_{j}\left( {\bf r}+z\widehat{{\bf z}}%
\right) \Psi _{j}^{\dagger }\left( {\bf r},z\right) \Psi _{j}\left( {\bf r}%
,z\right) .
\end{equation}
In this expression, ${\bf r}$ is a vector in the plane of the 2D gas. The
Fourier transformed polarisation operator is given by 
\begin{eqnarray}
{\bf p}\left( {\bf q},q_{z}\right) &=&\int d{\bf r}e^{-i{\bf q}_{\bot }\cdot 
{\bf r}}\int dze^{-iq_{z}z}{\bf p}\left( {\bf r},z\right)  \label{polanew} \\
&=&-e\sum_{j}\left[ \widehat{{\bf z}}\Gamma _{1,j}\left( q_{z}\right)
n_{j}\left( {\bf q}\right) +i\Gamma _{0,j}\left( q_{z}\right) \left[ {\bf %
\nabla }_{{\bf q}}n_{j}\left( {\bf q}\right) \right] \right] ,  \nonumber
\end{eqnarray}
where the matrix elements $\Gamma _{m,,j}\left( q_{z}\right) $ with $m=0,1$
are given by 
\begin{equation}
\Gamma _{m,,j}\left( q_{z}\right) =\int dze^{-iq_{z}z}\chi _{j}^{\ast
}\left( z\right) z^{m}\chi _{j}\left( z\right) ,
\end{equation}
and ${\bf q}$ is now redefined as a vector in the plane of the 2DEG. $%
n_{j}\left( {\bf q}\right) $ is the Fourier-transformed density operator for
well $j$. For propagation of the wave along the growth axis or in the plane
of the 2DEG, we can assume that $q_{z}\approx 0$. (In the later case, the
period of the SLS\ can always be made much smaller than the wavelength of
the light wave by appropriately tilting the sample {\it i.e.} by avoiding
the region too close to the C-SLS transition). Then, 
\begin{equation}
{\bf p}\left( {\bf q}\right) \equiv {\bf p}\left( {\bf q},q_{z}=0\right) =-ed%
\widehat{{\bf z}}\Lambda \left( {\bf q}\right) -ie\overrightarrow{\Theta }%
\left( {\bf q}\right) ,
\end{equation}
where 
\begin{eqnarray}
\Lambda \left( {\bf q}\right) &=&\frac{n_{R}\left( {\bf q}\right)
-n_{L}\left( {\bf q}\right) }{2}, \\
\overrightarrow{{\bf \Theta }}\left( {\bf q}\right) &=&{\bf \nabla }_{{\bf q}%
}\left[ n_{R}\left( {\bf q}\right) +n_{L}\left( {\bf q}\right) \right] .
\end{eqnarray}
The operator $\Lambda \left( {\bf q}\right) $ is related to fluctuations of
the relative electronic populations in the two wells while the operator $%
\Theta \left( {\bf q}\right) $ is related to fluctuations in the total
density of electrons.

The retarded polarisation response function of the inhomogeneous state is
defined, at $T=0$ K by 
\begin{equation}
\widetilde{{\bf \chi }}\left( {\bf q},{\bf q}^{\prime },\omega \right) =-%
\frac{i}{\hbar }\left[ \left\langle \left[ {\bf p}\left( {\bf q},t\right) ,%
{\bf p}\left( -{\bf q}^{\prime },t^{\prime }\right) \right] \right\rangle
\theta \left( t-t^{\prime }\right) \right] _{\omega },  \label{huit}
\end{equation}%
where $\left[ ... \right] _{\omega }$ stands for a Fourier transform in
time. In terms of the operators $\Lambda $ and $\overrightarrow{\Theta },$
we have 
\begin{eqnarray}
\widetilde{{\bf \chi }}\left( {\bf q},{\bf q}^{\prime },\omega \right) 
&=&e^{2}d^{2}\widehat{{\bf z}}\widehat{{\bf z}}\chi _{\Lambda ,\Lambda
}\left( {\bf q},{\bf q}^{\prime },\omega \right) +ie^{2}d\widehat{{\bf z}}%
\chi _{\Lambda ,{\bf \Theta }}\left( {\bf q},{\bf q}^{\prime },\omega
\right)  \\
&&+ie^{2}d\chi _{{\bf \Theta },\Lambda }\left( {\bf q},{\bf q}^{\prime
},\omega \right) \widehat{{\bf z}}-e^{2}{\bf \chi }_{{\bf \Theta },{\bf %
\Theta }}\left( {\bf q},{\bf q}^{\prime },\omega \right) ,  \nonumber
\end{eqnarray}%
and we can finally write for the absorption in the modulated coherent
states: 
\begin{eqnarray}
P\left( \omega \right)  &=&\frac{e^{2}\omega \left| E_{0}\right| ^{2}}{2V^{2}%
}\lim_{{\bf q}\rightarrow 0}%
%TCIMACRO{\func{Im}}%
%BeginExpansion
\mathop{\rm Im}%
%EndExpansion
\left[ d^{2}\,\left( {\bf \xi }\cdot \widehat{{\bf z}}\right) ^{2}\chi
_{\Lambda ,\Lambda }\left( {\bf q},{\bf q},\omega \right) \right] 
\label{neuf} \\
&&-\frac{e^{2}\omega \left| E_{0}\right| ^{2}}{2V^{2}}\lim_{{\bf q}%
\rightarrow 0}%
%TCIMACRO{\func{Im}}%
%BeginExpansion
\mathop{\rm Im}%
%EndExpansion
\left[ {\bf \xi \cdot \chi }_{{\bf \Theta },{\bf \Theta }}\left( {\bf q},%
{\bf q},\omega \right) \cdot {\bf \xi }\right]   \nonumber \\
&&+\frac{e^{2}d\omega \left| E_{0}\right| ^{2}}{2V^{2}}\lim_{{\bf q}%
\rightarrow 0}%
%TCIMACRO{\func{Re}}%
%BeginExpansion
\mathop{\rm Re}%
%EndExpansion
\left[ \left( {\bf \xi }\cdot \widehat{{\bf z}}\right) \left( \chi _{\Lambda
,{\bf \Theta }}\left( {\bf q},{\bf q},\omega \right) \cdot {\bf \xi }\right) %
\right]   \nonumber \\
&&+\frac{e^{2}d\omega \left| E_{0}\right| ^{2}}{2V^{2}}\lim_{{\bf q}%
\rightarrow 0}%
%TCIMACRO{\func{Re}}%
%BeginExpansion
\mathop{\rm Re}%
%EndExpansion
\left[ \left( {\bf \xi \cdot }\chi _{{\bf \Theta },\Lambda }\left( {\bf q},%
{\bf q},\omega \right) \right) \left( {\bf \xi }\cdot \widehat{{\bf z}}%
\right) \right] .  \nonumber
\end{eqnarray}%
More specifically, this expression gives the absorption for an incident
electromagnetic wave linearly polarised along ${\bf \xi }$ and propagating
in the direction ${\bf Q}=\left( {\bf q},q_{z}\right) $.

Exception made of the first term in Eq. (\ref{neuf}), all terms involve the
scalar product of the polarization vector with a vector in the plane of the
2D gas. The calculation of the absorption for an arbitrary propagation
direction of the incoming wave is complicated because of the need to solve
for the response functions ${\bf \chi }_{{\bf \Theta },{\bf \Theta }},\chi
_{\Lambda ,{\bf \Theta }},\chi _{{\bf \Theta },\Lambda }$. To avoid these
complications, we will consider an experimental situation where the
electromagnetic wave propagates in the plane of the 2D gas with its
polarisation vector pointing in the $z$ direction. This imposes severe
restrictions to an absorption experiment because of the small area that is
covered by the DQWS! We believe, however, that our conclusions will not
change qualitatively if the light wave makes a small angle with respect to
the normal to the growth axis. In fact, as we showed in Ref. \cite{cotesls},
the pseudospin-charge coupling is extremely small in the SLS so that the
neglected term are probably very small. With ${\bf \xi }$ close to $\widehat{%
{\bf z}}$, we have 
\begin{equation}
P\left( \omega \right) =\frac{1}{2V^{2}}e^{2}\omega \left| E_{0}\right|
^{2}d^{2}\lim_{{\bf q}\rightarrow 0}%
%TCIMACRO{\func{Im}}%
%BeginExpansion
\mathop{\rm Im}%
%EndExpansion
\left[ \chi _{\Lambda ,\Lambda }\left( {\bf q},{\bf q},\omega \right) \right]
.\qquad {\bf \xi }\Vert \widehat{{\bf z}}  \label{abso2}
\end{equation}
For normal incidence, ${\bf \xi }\bot \widehat{{\bf z}}$, the absorption is
related to the density correlation function only (more precisely, the gradient
with respect to wavevector of the density which is also the polarisation 
function in the plane of the wells).

In the extreme quantum limit where only one Landau level is occupied, it is
convenient to characterize the various ground states by the average value of
the operator $\rho _{i,j}\left( {\bf q}\right) $ defined by 
\begin{equation}
\rho _{i,j}\left( {\bf q}\right) =\frac{1}{N_{\phi }}%
\sum_{X}e^{-iq_{x}X-iq_{x}q_{y}\ell ^{2}/2}c_{i,X}^{\dagger
}c_{j,X+q_{y}\ell ^{2}}.
\end{equation}
The diagonal elements of this operator are related to the density of
electrons in the right ($\rho _{R,R}\left( {\bf q}\right) $) or left well ($%
\rho _{L,L}\left( {\bf q}\right) $). The off-diagonal terms, $\rho
_{R,L}\left( {\bf q}\right) ,\rho _{L,R}\left( {\bf q}\right) $ describe
coherence between the two wells. If the separation between the wells is
smaller than some critical value, it is possible for these coherence terms
to be non-zero even if the tunneling term itself is zero.

In the pseudospin representation, the total density and pseudospin density
operators are given by 
\begin{eqnarray}
\rho \left( {\bf q}\right) &=&\rho _{RR}\left( {\bf q}\right) +\rho
_{LL}\left( {\bf q}\right) , \\
S_{z}\left( {\bf q}\right) &=&\frac{1}{2}\left[ \rho _{RR}\left( {\bf q}%
\right) -\rho _{LL}\left( {\bf q}\right) \right] .
\end{eqnarray}
Any superposition of the $\left| R\right\rangle $ and $\left| L\right\rangle 
$ states can be mapped into an eigenstate of the pseudospin operator. In
particular, the operator $\rho _{R,L}$ and $\rho _{L,R}$ can be mapped in to
the pseudospin raising, $S_{+}$, and lowering, $S_{-}$, operators 
\begin{eqnarray}
S_{+}\left( {\bf q}\right) &=&S_{x}\left( {\bf q}\right) +iS_{y}\left( {\bf q%
}\right) =\rho _{RL}\left( {\bf q}\right) , \\
S_{-}\left( {\bf q}\right) &=&S_{x}\left( {\bf q}\right) -iS_{y}\left( {\bf q%
}\right) =\rho _{LR}\left( {\bf q}\right) .
\end{eqnarray}
In the pseudospin representation, the z-component of the polarisation
operator takes the form 
\begin{equation}
p_{z}\left( {\bf q}\right) \equiv -edN_{\phi }e^{-q^{2}\ell
^{2}/4}S_{z}\left( {\bf q}\right) .  \label{polarisation}
\end{equation}
If we keep only the fluctuations in $S_{z}$, then the absorption is given by 
\begin{equation}
P\left( \omega \right) =\frac{N_{\phi }}{2V^{2}}e^{2}\omega \left|
E_{0}\right| ^{2}d^{2}\lim_{{\bf q}\rightarrow 0}%
%TCIMACRO{\func{Im}}%
%BeginExpansion
\mathop{\rm Im}%
%EndExpansion
\left[ \chi _{S_{z},S_{z}}\left( {\bf q},{\bf q},\omega \right) \right]
,\qquad {\bf \xi }\Vert \widehat{{\bf z}}  \label{abso3}
\end{equation}
where the pseudospin response functions $\chi _{S_{z},S_{z}}$ is computed
from an analytic continuation of the finite temperature, Matsubara
two-particle Green's function 
\begin{equation}
\Gamma _{S_{z},S_{z}}\left( {\bf q},\Omega _{n}\right) =-N_{\phi
}\left\langle T\delta S_{z}\left( {\bf q},\tau \right) \delta S_{z}\left( -%
{\bf q},0\right) \right\rangle _{\Omega _{n}},
\end{equation}
where $\delta S_{z}{\bf \equiv }S_{z}{\bf -}\left\langle S_{z}\right\rangle $
and $\Omega _{n}$ is a Matsubara bosonic frequency.

\section{Commensurate-incommensurate transition in a DQWS}

In the pseudospin description, the Hartree-Fock energy per particle for the
2DEG in a DQWS subjected to an in-plane magnetic field can be written, at $%
\nu =1,$ as 
\begin{eqnarray}
E_{HF} &=&\frac{d}{l_{\bot }}\left\langle S_{z}\left( 0\right) \right\rangle
^{2}  \label{hartree} \\
&&-2\widetilde{t}%
%TCIMACRO{\func{Re}}%
%BeginExpansion
\mathop{\rm Re}%
%EndExpansion
\left[ \left\langle S_{+}\left( {\bf Q}\right) \right\rangle \right] 
\nonumber \\
&&+\frac{1}{4}\sum_{{\bf q}}\Upsilon \left( {\bf q}\right) \left\langle \rho
\left( -{\bf q}\right) \right\rangle \left\langle \rho \left( {\bf q}\right)
\right\rangle  \nonumber \\
&&+\sum_{{\bf q}}J_{z}\left( {\bf q}\right) \left\langle S_{z}\left( -{\bf q}%
\right) \right\rangle \left\langle S_{z}\left( {\bf q}\right) \right\rangle 
\nonumber \\
&&+\sum_{{\bf q}}J_{\bot }\left( {\bf q}\right) \left\langle {\bf S}_{\bot
}\left( -{\bf q}\right) \right\rangle \cdot \left\langle {\bf S}_{\bot
}\left( {\bf q}\right) \right\rangle .  \nonumber
\end{eqnarray}
All energies in this equation are in units of $e^{2}/\kappa l_{\bot }$. The
tunneling amplitude is given by 
\begin{equation}
\widetilde{t}=te^{-d^{2}\ell ^{2}/4\ell _{||}^{4}}\equiv te^{-Q^{2}\ell
^{2}/4}.
\end{equation}
In the gauge we are using, the parallel component of the magnetic field is
responsible for the introduction of a guiding-center-dependent phase factor
that depends on the $X$ quantum number only {\it i.e.} $e^{-iXd/\ell
_{||}^{2}}\equiv e^{-iQX}$ where $Q$ is defined by 
\begin{equation}
Q\equiv \frac{d}{\ell _{||}^{2}}\propto B_{||}.
\end{equation}
This is why $\left\langle S_{+}\left( {\bf Q}\right) \right\rangle $ appears
in Eq.(\ref{hartree}), instead of $\left\langle S_{+}\left( {\bf 0}\right)
\right\rangle .$ In Eq. (\ref{hartree}), we have defined 
\begin{equation}
J_{z}\left( {\bf q}\right) =V_{a}\left( {\bf q}\right) -V_{b}\left( {\bf q}%
\right) -V_{c}\left( {\bf q}\right) ,
\end{equation}
and 
\begin{equation}
\Upsilon \left( {\bf q}\right) =V_{a}\left( {\bf q}\right) 
-V_{b}\left( {\bf q}\right) 
+V_{c}\left( {\bf q}\right) ,
\end{equation}
and 
\begin{equation}
J_{\bot }\left( {\bf q}\right) =-V_{d}\left( {\bf q}\right) .
\end{equation}
where $V_{a}$ and $V_{c}$ are the Hartree intra and inter-well interactions
and $V_{b}$ and $V_{c}$ are the exchange (Fock) intra and inter-well
interactions. These interactions are defined by 
\begin{eqnarray}
V_{a}({\bf q}) &=&\left( \frac{1}{q\ell }\right) e^{-q^{2}\ell ^{2}/2}, \\
V_{b}({\bf q}) &=&\int_{0}^{\infty }d(q^{\prime }\ell )J_{0}(qq^{\prime
}\ell ^{2})e^{-q^{\prime 2}\ell ^{2}/2}, \\
V_{c}({\bf q}) &=&\left( \frac{1}{q\ell }\right) e^{-q^{2}\ell
^{2}/2}e^{-qd}, \\
V_{d}({\bf q}) &=&\int_{0}^{\infty }d(q^{\prime }\ell )J_{0}(qq^{\prime
}\ell ^{2})e^{-q^{\prime 2}\ell ^{2}/2}e^{-qd}.
\end{eqnarray}

We now give a brief summary of the commensurate-incommensurate transition.
In this transition, the DQWS is kept at $\nu =1$ while the sample is
inclined by an angle $\alpha $. The commensurate phase that appears when $%
\alpha <\alpha _{c}$ is described by the order parameters 
\begin{eqnarray}
\left\{ \left\langle S_{z}\left( {\bf q}\right) \right\rangle \right\}
&=&0,\left\{ \rho \left\langle \left( {\bf q}\neq 0\right) \right\rangle
\right\} =0,\rho \left\langle \left( 0\right) \right\rangle =1, \\
&<&S_{+}(Q,0)>=\frac{1}{2}.
\end{eqnarray}
The last equation indicates that 
\begin{equation}
\left\langle {S}_{+}\left( X\right) \right\rangle =<c_{R,X}^{\dagger
}c_{L,X}>=\frac{1}{2}e^{i\varphi (X)},
\end{equation}
with $\varphi (X)=QX.$ Hence, the pseudospin rotates in space according to $%
\left( \left\langle S_{x}\left( X\right) \right\rangle ,\left\langle
S_{y}\left( X\right) \right\rangle \right) =\left( \cos \left( QX\right)
,\sin \left( QX\right) \right) $. At this point, the analysis is simplified
if we define a new phase $\widetilde{\varphi }(X)$ by 
\begin{equation}
\widetilde{\varphi }(X)=\varphi (X)-QX,  \label{aquinze}
\end{equation}
so that in the commensurate state, $\widetilde{\varphi }(X)\equiv 0.$ If we
use a tilde to denote the operators in the ``rotating frame'', we have 
\begin{equation}
<\widetilde{S}_{+}(q_{x})>\equiv <S_{+}(q_{x}+Q)>=\frac{1}{2N_{\phi }}%
\sum_{X}e^{-iq_{x}X}e^{i\widetilde{\varphi }(X)},
\end{equation}
and so in the commensurate phase 
\begin{equation}
<\widetilde{S}_{+}(q_{x})>=<\widetilde{S}_{x}(q_{x})>=\frac{1}{2}\delta
_{q_{x},0}.
\end{equation}

When the magnetic field is tilted above a critical angle, the energy to
create defects in the form of solitons becomes negative. These solitons
corresponds to slips of $2\pi $ in the pseudospin texture. In the rotating
frame and in the so-called gradient approximation summarized in Ref.\cite%
{cotesls}, they are kinks given by 
\begin{equation}
\widetilde{\varphi }(X)=4\tan ^{-1}\left[ e^{-\sqrt{\frac{\widetilde{t}}{%
2\pi \rho _{s}\ell ^{2}}}X}\right] ,  \label{bsept}
\end{equation}
where $\rho _{s}$ is the spin stifness of the system (which is basically due
to the Fock inter-well interaction given above). Because of the repulsive
interaction between solitons, there is, at each value of the parallel
magnetic field, an optimal density of solitons. In the ground state, these
solitons condense into a crystal with a period $L_{s}$ so that $e^{i%
\widetilde{\varphi }(X+L_{S})}=e^{i\widetilde{\varphi }(X)}.$ This soliton
lattice is described by the set of order parameters 
\begin{gather}
\left\{ \left\langle S_{z}\left( {\bf q}\right) \right\rangle \right\} =0, \\
\left\{ \rho \left\langle \left( {\bf q}\neq 0\right) \right\rangle \right\}
=0,\rho \left\langle \left( 0\right) \right\rangle =1, \\
<\widetilde{S}_{+}(q_{x}=nQ_{s},q_{y}=0)>\neq 0,
\end{gather}
where 
\begin{equation}
Q_{s}=\frac{2\pi }{L_{s}}  \label{qsoliton}
\end{equation}
is the wavevector of the soliton lattice and is a function of the parallel
component of the magnetic field.

The energy of the commensurate state is 
\begin{equation}
E_{C}=-\widetilde{t}-\frac{1}{4}\left( \frac{e^{2}}{\kappa \ell }\right)
V_{d}(Q).  \label{enercommensurable}
\end{equation}
It increases monotonically with the magnetic field. In the limit of strong
parallel magnetic fields, the energy of the soliton-lattice state becomes
equal to the energy of an incommensurate state described by $\varphi (X)=0$ 
{\it i.e.}, $\widetilde{\varphi }(X)=-QX$ or, equivalently, $<\widetilde{%
\rho }_{RL}(-{\bf Q)}>=\frac{1}{2}$ (all other parameters being zero). In
this limit, $Q_{s}\rightarrow Q$ so that the solitons are spaced by $2\pi
/Q. $The energy of the I state is given by 
\begin{equation}
E_{I}=-\frac{1}{4}\left( \frac{e^{2}}{\epsilon _{0}\ell }\right) V_{d}(0),
\label{enerincommensurable}
\end{equation}
and is clearly independent of the parallel magnetic field and of the
tunneling term. These energies are plotted in Fig. 1.

\section{Absorption in the commensurate and incommensurate states}

In Ref. \cite{cotesls}, we computed the density and pseudospin response
functions in the C,I and SL states in the time-dependent Hartree-Fock
approximation (TDHFA). In the commensurate state, an analytical solution for
the pseudospin response is 
\begin{equation}
\chi _{S_{z},S_{z}}\left( {\bf q},{\bf q}^{\prime },\omega \right) =-\left( 
\frac{b\left( {\bf q}\right) }{2}\right) \frac{\left[ 1+\cos \left(
Qq_{y}\ell ^{2}/2\right) \right] \delta _{{\bf q},{\bf q}^{\prime }}}{\left(
\omega +i\delta \right) ^{2}-\omega _{C}^{2}\left( {\bf q}\right) },
\end{equation}
where the frequency of the collective mode is given by 
\begin{equation}
\omega _{C}^{2}\left( {\bf q}\right) =4a\left( {\bf q}\right) b\left( {\bf q}%
\right) ,
\end{equation}
with 
\begin{equation}
a\left( {\bf q}\right) =t_{R}+\frac{1}{2}\left[ V_{a}\left( {\bf q}\right)
-V_{b}\left( {\bf q}\right) -V_{c}\left( {\bf q}\right) \cos \left(
Qq_{y}\ell ^{2}/2\right) \right] ,
\end{equation}
\begin{equation}
b\left( {\bf q}\right) =t_{R}-\frac{1}{4}\left[ V_{d}\left( {\bf q+}Q%
\widehat{{\bf x}}\right) +V_{d}\left( {\bf q-}Q\widehat{{\bf x}}\right) %
\right] ,
\end{equation}
and 
\begin{equation}
t_{R}=\widetilde{t}+\frac{1}{2}V_{d}\left( Q\right) .
\end{equation}
In the limit ${\bf q}\rightarrow 0,$ the absorption is then proportionnal to 
\begin{eqnarray}
P_{C}\left( \omega \right) &\sim &\omega \left| E_{0}\right| ^{2}d^{2}\lim_{%
{\bf q}\rightarrow 0}%
%TCIMACRO{\func{Im}}%
%BeginExpansion
\mathop{\rm Im}%
%EndExpansion
\left[ \chi _{S_{z},S_{z}}\left( {\bf q},{\bf q},\omega \right) \right]
\qquad {\bf \xi }\Vert \widehat{{\bf z}} \\
&\sim &\left| E_{0}\right| ^{2}d^{2}\widetilde{t}\delta \left( \omega
-\omega _{C}\left( {\bf q}=0\right) \right)
\end{eqnarray}
where 
\begin{equation}
\omega _{C}\left( {\bf q}=0\right) =2\sqrt{\widetilde{t}\left( \widetilde{t}+%
\frac{d}{2\ell }-\frac{1}{2}\sqrt{\frac{\pi }{2}}+\frac{1}{2}V_{d}\left(
Q\right) \right) }  \label{renorm}
\end{equation}
gives the gap in the dispersion relation of the pseudospin wave in the C
phase. This gap is shifted from it's noninteracting value, $2\widetilde{t}$
because of many-body exchange and vertex corrections. In the absence of
vertex corrections ({\it i.e.} in the Hartree-Fock approximation), the SAS
gap is renormalized to $2t_{R}.$ The vertex corrections produce a
substantial reduction of the Hartree-Fock gap.

From the expression of $P_{C}\left( \omega \right) $, we see that there is
no absorption in the absence of tunneling when the pseudospin wave mode
given by Eq. (\ref{renorm}) is gapless. The absorption is non zero, however,
in the absence of the parallel magnetic field, if $t\neq 0$ and for ${\bf %
\xi }\Vert \widehat{{\bf z}}.$ 

In the I state, we have 
\begin{equation}
\chi _{S_{z},S_{z}}\left( {\bf q},{\bf q}^{\prime },\omega \right) =\frac{%
\left[ V_{d}\left( {\bf q}\right) -V_{d}\left( 0\right) \right] \delta _{%
{\bf q},{\bf q}^{\prime }}}{\left( \omega +i\delta \right) ^{2}-\omega
_{I}^{2}\left( {\bf q}\right) },
\end{equation}%
where 
\begin{eqnarray}
\omega _{I}\left( {\bf q}\right)  &=&\sqrt{\left( V_{d}\left( 0\right)
-V_{d}\left( {\bf q}\right) \right) }  \label{omegi} \\
&&\times \sqrt{\left( V_{a}\left( {\bf q}\right) -V_{b}\left( {\bf q}\right)
-V_{c}\left( {\bf q}\right) +V_{d}\left( {\bf q}\right) \right) }.  \nonumber
\end{eqnarray}%
There is no signal in the absorption spectrum in this case: 
\begin{equation}
P_{I}\left( \omega \right) =0.\qquad {\bf \xi }\Vert \widehat{{\bf z}}
\end{equation}

\section{Absorption in the soliton lattice state}

Fig. 1 shows the energy of the C, I and SL states calculated in the
Hartree-Fock approximation with the parameters $t/\left( e^{2}/\kappa \ell
\right) =0.01,d/\ell =1.0$ that we used for all the other results presented
in this work. The inset in Fig. 1 shows how the SL wavevector $Q_{s}$
evolves with the parallel magnetic field from the parallel magnetic field
wavevector $Q_{c}$ at the transition from the C to SL states. ($Q_{c}\approx
0.62$ for our choice of parameters). The SLS extents from approximately $%
Q=0.62$ to $Q=2.0$ where its energy is nearly indistinguishable from that of
the I state.

\begin{figure}[tbp]
\centerline{\epsfxsize 8cm \epsffile{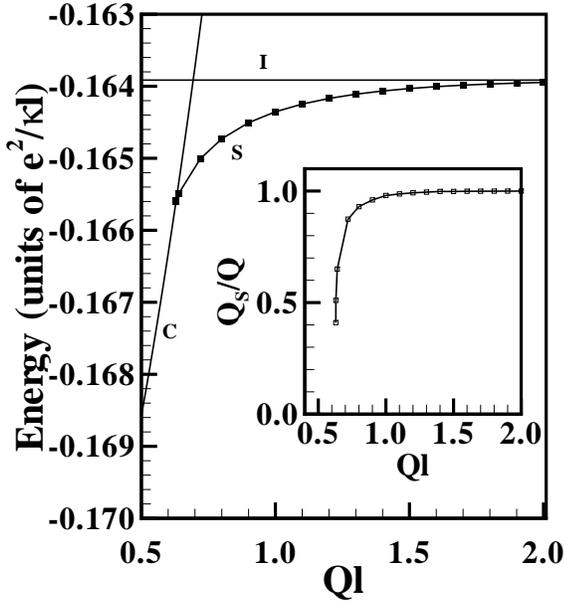}}
\caption{ Energy of the commensurate, incommensurate and soliton-lattice
states as a function of the parallel magnetic field. The inset shows the
dependence of the soliton-lattice wavevector on the parallel field.}
\label{fig1}
\end{figure}

It is not possible to solve analytically for the response functions in the
SLS. They must be obtained numerically. The procedure to obtain these
dispersions is explained in Ref. \cite{cotesls}. The SLS\ sustain many
branches of collective excitations that are all periodic along the $x$
direction (the parallel magnetic field is applied along $y$). Figs. 2 show
the low-energy part of the the dispersion in the first Brillouin zone of the
SLS for wavevectors $Q\ell =0.64,1.0,2.0$. When the parallel field is strong
(Fig. 2(c)), the branches of collective excitations are exactly given by $%
\omega _{I}\left( {\bf q}={\bf k+}nQ_{s}\widehat{{\bf x}}\right) $ where $%
n=0,\pm 1,\pm 2,... $, and ${\bf k}$ is a vector restricted to the first
Brillouin defined of the SL. In this high-parallel field limit, the
pseudospin modulations in the SLS results in a folding of the collective
modes of the $t=0$ ground state inside the first Brillouin zone. As the
parallel field is decreased, the coupling between different modes increases
and gaps open up in the dispersion. As $Q\rightarrow Q_{c}$, the dispersion
becomes increasingly different from $\omega _{I}\left( {\bf k}+nQ_{s}%
\widehat{{\bf x}}\right) $.

\begin{figure}[tbp]
\centerline{\epsfxsize 8cm \epsffile{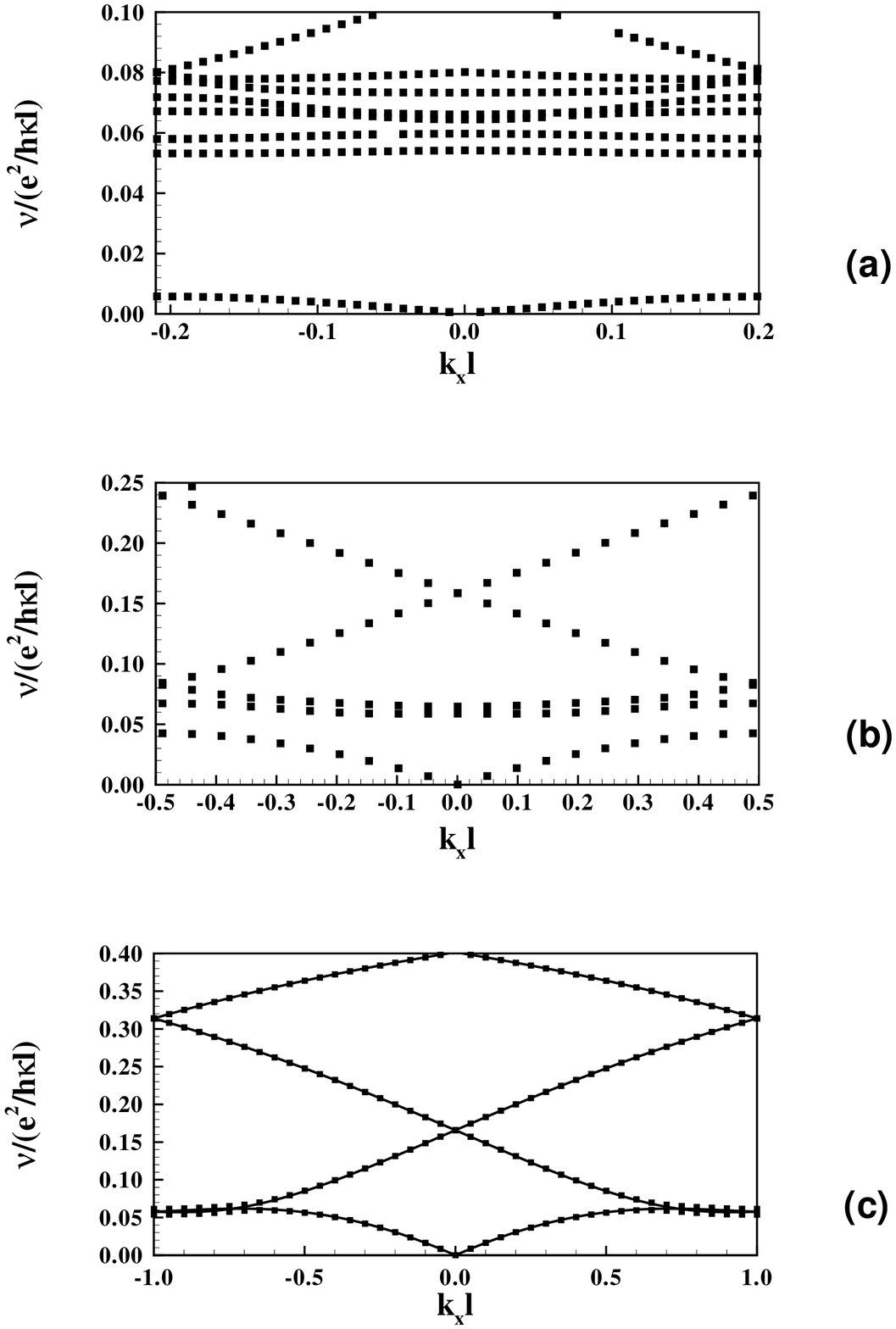}}
\caption{ Low-energy dispersion branches of the collective modes in the SLS
for $t/(e^{2}/k\ell _{\bot })=0.01,$ $d/\ell _{\bot }=1.0$ and (a) $Q\ell
_{\bot }=0.64$, (b) $Q\ell _{\bot }=1.0$, (c) $Q\ell _{\bot }=2.0.$ The full
lines in (c) are given by 
$
\omega _{I}\left( {\bf q}={\bf k+}nQ_{s}\widehat{{\bf x}}\right) $ with $%
n=0,\pm 1,\pm 2.$
}
\label{fig2}
\end{figure}

When the parallel magnetic field is large, the folding of the modes inside
the Brillouin zone results in a set $\left\{ \omega _{n}\left( 0\right)
\right\} $ of ${\bf k}=0$ modes that correspond to the frequencies $\left\{
\omega _{I}\left( nQ_{s}\widehat{{\bf x}}\right) \right\} .$ As the parallel
field decreases, these frequencies evolves into a serie of curves
represented in Fig. 3. This set of curves is a distinctive feature of the
SLS and would be a good signature of its existence. Unfortunately, the
absorption spectrum does not capture all of these excitations. In fact, very
few branches survive in $P\left( \omega \right) $ as is clear from Fig. 4.
The absorption spectrum consists of a broad peak near the transition at $%
Q_{c}$ (at a frequency close to the renormalized tunneling energy) that
further spreads into a number of well-defined peaks as the parallel magnetic
field is increased. Very rapidly, however, only two of the peaks
(corresponding to the lowest two branches in Fig. 3) survive. At larger
parallel field, only the lowest-energy branch has significant weight in the
absorption spectrum. For still larger fields, when the system asymptotically
approaches the incommensurate state, the absorption disappears completely.
Below the transition to the SLS, the absorption spectrum consists of only
one peak at the renormalized value of the gap energy as we have shown above.
Fig. 4 shows that there is a definite signature of the SLS in the absorption
spectrum although it is not as pronounced as we might first have expected.

\begin{figure}[tbp]
\centerline{\epsfxsize 8cm \epsffile{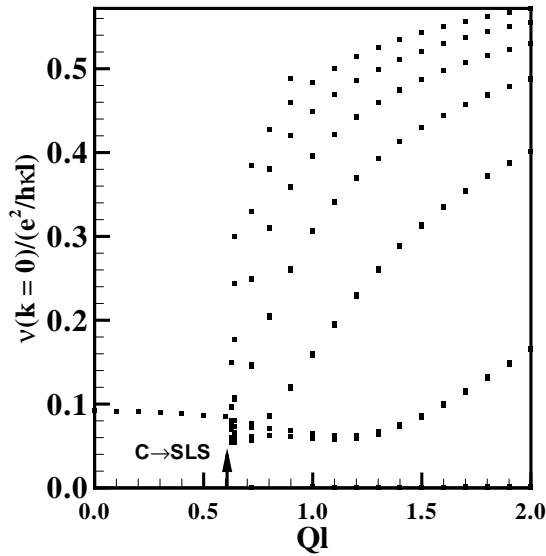}}
\caption{Evolution of the ${\bf {k}=0}$ frequencies of the collective modes
branches from the commensurate to the SL state.}
\label{fig3}
\end{figure}

\begin{figure}[tbp]
\centerline{\epsfxsize 8cm \epsffile{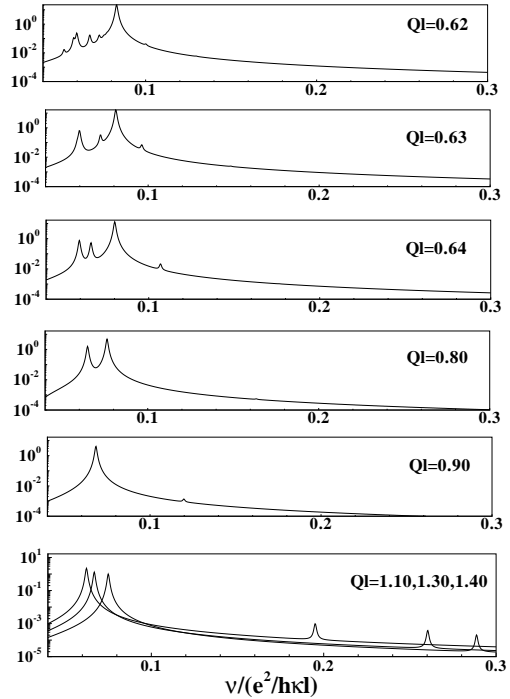}}
\caption{Absorption spectrum (in arbitrary units) of the soliton-lattice
state for several values of the parallel magnetic field.}
\label{fig4}
\end{figure}

From Fig. 4, we also see that the large peak in the absorption spectrum
occurs near the renormalized gap energy given approximately by $\omega
_{C}\left( 0\right) $ of Eq. (\ref{renorm}). A large fraction of this gap
comes from many-body corrections. For parameters appropriate to the
weak-tunneling sample of Murphy et al.\cite{murphy},{\it \ i.e.} $%
n_{s}=1.26\times 10^{11}$ cm$^{-2}$, $d=211$ \AA\ with a small tunneling
energy of $t=0.4$ K, we have $d/\ell =1.87$ and $t/\left( e^{2}/\kappa
\ell \right) =0.003$ so that $\hbar \omega _{C}\left( 0\right) \approx
0.1\left( e^{2}/\kappa \ell \right) $ just as for the parameters we choose
in this paper. An energy of $\hbar \omega =0.1$ $\left( e^{2}/\kappa \ell
\right) $ corresponds to a frequency of approximately $2.6\times 10^{11}$ Hz
(if it were observable in the absorption, the highest-energy branch would
correspond to a maximal frequency of approximately $1.6\times 10^{12}$ Hz).
This places the interesting features in the absorption spectrum in the
far-infrared region of the electromagnetic spectrum, a difficult region to
investigate with available laser sources. It is possible to increase $\omega
_{C}\left( 0\right) $ by a factor of $5$ or more using a DQWS with a
stronger tunneling gap or by modifying the other parameters of the sample
(provided this choice of parameters does not place the sample outside the
region of stability of the coherent state) but observation would still
remain difficult.

\section{Conclusion}

We have computed the absorption spectrum of the 2DEG in a DQWS when the
sample is gradually tilted with respect to the quantizing magnetic field. At
filling factor $\nu =1$, the commensurate-incommensurate transition driven
by the parallel field is reflected in a change of behaviour of the
absorption spectrum. The soliton lattice that is the ground state of the
2DEG above the transition has a distinctive set of collective excitations.
Some of these excitations can be seen, in principle, in the absorption
spectrum in a small region above the C-I transition. Experimental
observation of this behaviour, however, is expected to be difficult.

\section{Acknowledgments}

This research was supported by a grant from the Natural Sciences and
Engineering Research Council of Canada (NSERC). The author want to thank S.
Charlebois and J. Beerens for helpful discussions.

\end{document}